\documentclass[prl,aps,twocolumn,amsfonts,showpacs,superscriptaddress]{revtex4-1}

\newsavebox{\foobox}

\usepackage{isomath}
\usepackage{graphicx}
\usepackage{xcolor}
\usepackage{rotating}
\usepackage{amsmath,amssymb,graphics}

\usepackage[colorlinks=true, urlcolor=violet, linkcolor=blue, citecolor=red, hyperindex=true, linktocpage=true]{hyperref}

\def\ben{\begin{equation}}
\def\een{\end{equation}}

\def\be{\begin{equation}}
\def\ee{\end{equation}}
\def\beq{\begin{equation}}
\def\eeq{\end{equation}}
\def\ba{\begin{array}}
\def\ea{\end{array}}

\def\dalemb#1#2{{\vbox{\hrule height .#2pt
\hbox{\vrule width.#2pt height#1pt \kern#1pt
\vrule width.#2pt}
\hrule height.#2pt}}}

\newcommand{\bea}{\begin{eqnarray}}
\newcommand{\eea}{\end{eqnarray}}
\newcommand{\mt}[1]{\textrm{\tiny #1}}

\newcommand{\rh}{r_\mt{H}}

\newcommand{\me}{\mathrm{e}}
\newcommand{\md}{\mathrm{d}}

\begin{document}

\title{Universality of DC Electrical Conductivity from Holography}

\author{Xian-Hui Ge}
\email{gexh@shu.edu.cn}
\affiliation{ Shanghai Key Laboratory of High Temperature Superconductors, Department of Physics, Shanghai University, Shanghai, 200444, China \\ and Department of Physics, University of California at San Diego, CA92122, USA}

\author{Sang-Jin Sin}
\email{sjsin@hangyang.ac.kr}

\affiliation{Department of Physics, Hanyang University, Seoul 133-791, Korea}

\author{Shao-Feng Wu}
\email{sfwu@shu.edu.cn}

\affiliation{Department of Physics, Shanghai University, Shanghai, 200444, China}

\date{\today}

\begin{abstract}
We propose a universal formula of  dc electrical conductivity in rotational- and translational- symmetries breaking systems via the holographic duality.
This formula states that the ratio of the determinant of the dc electrical conductivities along any spatial directions to the black hole area density in zero-charge limit has a universal value. As explicit illustrations, we give several examples elucidating the validation of this formula: We construct an anisotropic black brane solution, which yields linear in temperature for the in-plane resistivity and insulating behavior for
the out-of-plane resistivity; We also construct a spatially isotropic black brane solution that both the linear-T and
quadratic-T contributions to the resistivity can be realized.
\end{abstract}

\maketitle

\section{1. Introduction}
The AdS/CFT correspondence provides a powerful tool  to analyze  strongly coupled systems, particularly for studying the transport properties of strongly coupled systems. One of the most famous results
of the AdS/CFT applications, is the so-called Kovtun-Son-Starinets (KSS) bound $\eta/s\geq \hbar/(4 \pi k_B)$, which states that for strongly coupled systems with a classical Einstein gravity dual description, the ratio of the shear
viscosity $\eta$, to the entropy density $s$, obeys such a bound \cite{kss}. In most higher derivative gravity models, the bound is violated and there may still exist a lower bound \cite{Cremonini:2011iq,ge08,liu08,ge09,cai,myers}, but even this is not clear \cite{buchel}.

Recently, the KSS conjecture was severely challenged by the anisotropic black brane systems, where  the shear viscosity is a tensor and some components of the tensor can become considerably smaller, which parametrically violates the bound \cite{rehban,mamo}. Considered a $d+1$-dimensional geometry with coordinates $(t,x_i,z)$,  and anisotropy only along the $z-$direction, the shear viscosity to entropy density ratio is related to the anisotropy as follows\be
\frac{\eta_{x_i z,x_i z}}{s}=\frac{\hbar}{4\pi k_B} \frac{g_{x_i x_i}}{g_{zz}}\bigg|_{r=\rh},
\ee
where $g_{x_i x_i}$ and $g_{zz}$ are the line elements of the metric and $\rh$ is the event horizon radius, respectively. For translational symmetry unbroken system, there is a universal relation between the graviton absorption cross--section and the black--brane horizon area in the large--incident--wavelength limit \cite{mathur}: $ \mathcal{A}= \Sigma(\omega=0)$.
 The rise of the event horizon area--graviton cross--section  equivalence is simply because the metric perturbation component $h_{x_i}^{~x^j}$ satisfies the equation of motion of the minimally coupled massless scalar $ \Box h_{x_i}^{~x^j}=0$. The spin-$2$ shear viscosity component  is proportional to the graviton absorption cross--section via $\eta_{x_i x_j,x_i x_j}={\Sigma(\omega=0)}/{2\kappa^2}$. Therefore, the spin-$2$ shear viscosity component is linearly dependent on the event horizon area (i.e. the entropy density). However, for the spin- 1 component $h_{z}^{x_i}$, the equation of motion is not identical to minimally coupled massless Klein-Gordon equation and thus  the absorption cross--section of spin--1 vector field $h_{x_i z}$
in an anisotropic black--brane background  is not equal to the black--brane horizon area.
An arbitrary violation of the KSS bound would occur if $g_{x_i x_i}/g_{zz}\rightarrow 0$. In  this anisotropic background, the rotational symmetry of the dual field theory is broken from $SO(d-1)$ to $SO(d-2)$.
We thus have shear viscosities $\eta_{x_i z,x_i z}$, which are defined by the metric fluctuations $h_{x_i z}$. Such metric components carry spin $1$ with respect to the $SO(d-2)$ symmetries \cite{jain}.  Although the spin- $2$ components of the shear viscosity tensor in the $x_i-x_j$ plane  satisfy the KSS bound, the shear force in the $x_i-z$ plane, which is related to the spin- $1$ metric components, can violates it.
Furthermore, the diffusion bound $D \gtrsim C { \hbar v^2_{F} }/{k_B T}$ (C is a constant) will also break down \cite{kovtun,hartnoll,pakhira1,pakhira2}. The diffusivity bound was proposed to replace the Mott-Ioffe-Regel (MIR) bound in bad metals, and it is based on the KSS bound $\eta/s \geq C {\hbar}/{k_{B}}$ and the relation ${\eta}/{s}={D T}/{c^2}$ for a vanishing chemical potential \cite{hartnoll}.

We also notice that even in isotropic systems, the KSS conjecture could be violated due to momentum dissipation \cite{wang,ling,hartnoll16,matteo,pj}. In translation invariance broken but isotropic systems, fluctuations of the metric components becomes massive and the corresponding shear viscosity does not yield a hydrodynamic description. In this case, the shear viscosity to entropy density ratio behaving as  $\eta/s \sim T^{\nu}$ with $\nu$ a positive parameter,  violating the KSS conjecture even in Einstein gravity. The shear viscosity now quantifies the rate of entropy production due to a strain.

One natural question is  whether there is an alternative bound to be obeyed by the transport coefficients in such anisotropic systems.
It is well-known that in condensed matter physics, it is notably universal that the materials are anisotropic with different properties in different directions. Remarkably, the transport in high--$T_c$ cuprates is strongly two-dimensional in character and there is  substantial anisotropy  between the in- and out-of-plane (i.e. $\rm CuO_2$ plane) resistivities. In contrast to the resistivity $\rho_{ab}$ in the $\rm CuO_2$ planes,   where a generic behavior is observed to  depend on the metallic temperature, the $c-$axis transport in high- temperature cuprates is very highly material-specific. Intriguingly, in most underdoped cuprates, $\rho_c(T)$ shows insulating behavior at all temperatures \cite{brooks}.

Therefore, the universal transporting properties of anisotropic systems deserve further studies. In this study, we will show that the ratio of the determinant of the electrical DC conductivities to the graviton absorption cross--section in anisotropic systems from holography in the zero--charges  limit has a universal value
\be \label{scaling}
\frac{\prod_{i} {\sigma_{ii}}}{ \mathcal{A}^{d-3}}\bigg|_{q_i=0} = \prod_{i} Z_i^{d-1}\bigg|_{r=\rh},
\ee
where $\mathcal{A}$ and $Z_i$ are  area density per unit volume of the black hole event horizon  and gauge field couplings, respectively. In the minimal coupling case, $Z_i=1$.
Isotropic systems can be considered as special case of anisotropic systems.


The universal relation  (\ref{scaling}) is able to provide us some insights into the holographic realizations of the linear temperature resistivity:\\
1). For $Z(\phi)=1$ and $d\geq 3$, isotropic black branes in the AdS space cannot be utilized to realize linear temperature resistivity in the zero-charges limit. Nevertheless, anisotropic black branes are good candidates
in model-building of holographic strange metals.\\
2). For $d+1$-dimensional spatially isotropic Lifshitz black holes with $Z(\phi)=1$ in the absence of hyperscaling violation, this relation indicates that $\sigma_{ii}|_{q_i=0}=[{4\pi}/(d+z-1)]^{d-3}T^{(d-3)/z}$, which is consistent with what obtained in Refs.\cite{damle,sachdev} based on a universal scaling relation hypothesis: $\sigma(\omega=0)=T^{(d-3)/z}\Theta(0)$, where $z$ is a dynamical critical exponent and  $\Theta(\omega)$ is a frequency dependent function.\\
3). This relation applies to shear viscosity-bound and electrical conductivity-bound violated systems, for example, systems considered in \cite{wang, weijia}. In \cite{schalm}, the authors conjectured that for the case
$d=3$, there exists a lower bound of dc electrical conductivity $\prod_{i} \sigma_{ii}> 1$.  But it was soon found that this bound can be violated by a special coupling between the linear axion fields and the $U(1)$ gauge field \cite{weijia}.

The structure of this paper is organized as follows. In section 2, we present our main results by writing down the conductivity tensor in terms of horizon data for anisotropic systems.
 We then present three examples that reproduce particular features of strange metals in section 3. Discussions and conclusions are presented in section 4

\section{2. Main Results}\label{sec:incoh}
Without loss of generality, we consider the Einstein-Maxwell-dilaton action with linear scalar fields
\bea\label{uniaction}
S&=&\int d^{d+1}x \sqrt{-g}\bigg[\frac{1}{16\pi G}\bigg(R-\frac{1}{2}\partial \phi^2+V(\phi)\nonumber\\&-&\frac{1}{2}\sum^{p-1}_{i=1}Y_i(\phi)\partial\psi^2_{i}\bigg)
-\frac{1}{4 g^2_{d+1}}Z(\phi)F^2\bigg].
\eea
Hereafter, we  select $16 \pi G=g^2_{d+1}=L=1$, where $L$ is the AdS radius, $g^2_{d+1}$ is the $d+1$-dimensional gauge coupling constant, and $G$ is  Newton's constant. Recently,  this model has been widely studied  in Refs. \cite{withers,kim,donos3,amoretti,amoretti1,cheng,jianpin,gauntlett}.
The solution to the above theory is assumed to be anisotropic
\bea
&&ds^2=-g_{tt}dt^2+g_{rr}dr^2+g_{xx}dx^jdx^j+g_{zz}dzdz,\\
&& \phi=\phi(r),~~~A=A_t(r)dt,~~\psi_{j}=k_j x_j, j=1\cdots d-2, \nonumber\\&&\psi_{z}=k_z z,  ~~k_j\neq k_z. \nonumber
\eea
The anisotropic direction is  selected along the $z$-direction. We regard the $x_i-x_j$ plane as the $``ab"$ plane and the $z-$direction as the $``c"-$axis in cuprates.
The entropy density is given by $s=4 \pi (g^{d-2}_{xx}g_{zz})^{1/2}|_{r=\rh}$. The electric charge density is given by $q\equiv -J^t=-\sqrt{-g}Z(\phi)\partial_r A_t$.

We impose a constant electric field in the $x_i$ direction with magnitude $E$, which will generate electric currents  only along the $x_j$ direction. Let us consider a small perturbation  in the black hole background
\bea
&&A_{j}=-E t+\delta a_{x_j} (r), ~~~g_{t x_j}=\delta g_{t x_j}(r),\nonumber\\&&~~~g_{r x_j}=g_{xx}\delta h_{r x_j}(r),~~~\psi=k_j x_j+\delta \chi_1.
\eea
From  Maxwell equation $\partial_r (\sqrt{-g} Z(\phi)F^{rx_i})=0$, we can define a conserved current $J^{x_j}=-\sqrt{-g}g^{rr}g^{xx}Z(\phi)\partial_r a_{x_j}+\delta g_{tx_j}g^{xx}q$.
In the absence of a charge density, we only have a contribution to the current from the gauge field $J^{x_j} \sim \partial_r a_{x_j}$. The conductivity can be determined  based on the horizon regularity. In this case, we simply haves
\be
\bigg(\sqrt{\frac{g_{tt}}{g_{rr}}}a'_{x_j}\bigg)'=0.
\ee
Regularity at the horizon gives us \be
a_{x_j}=-\frac{E}{4\pi T}\ln(r-\rh).
\ee
At finite charge density, we  must know the behavior of $\delta g_{tx_j}$ at the horizon. In the presence of  momentum dissipation, $\delta g_{tx_j}$ will take a finite value at the horizon
\be
\delta g_{tx_j}=\frac{Eq}{ k_j^2 Y_H g^{\frac{d-3}{2}}_{xx}}\bigg|_{r=\rh},
\ee
where we use the notation $Y_H=Y(\phi_H)$ and $Z_H=Z(\phi_H)$.
Therefore, the conserved current is obtained as
\be
J^{x_j}=\bigg(g^{\frac{d}{2}-2}_{xx}g^{\frac{1}{2}}_{zz}Z_H E+\frac{E q^2}{k^2_j Y_H g^{\frac{d-1}{2}}_{xx}}\bigg)\bigg|_{r=\rh}.
\ee
Then, the DC conductivity is  given by
\be
\label{isocon}
\sigma_{jj}=\frac{J^{x_j}}{E}=\bigg(g^{\frac{d}{2}-2}_{xx}g^{\frac{1}{2}}_{zz}Z_H +\frac{ q^2}{k^2_j Y_H g^{\frac{d-1}{2}}_{xx}}\bigg)\bigg|_{r=\rh}.
\ee
The first term in the above formula corresponds to the conductivity of the particle-hole pair creation and the second term is associated with the momentum relaxation.
In the following section, we will calculate the
DC conductivity along the anisotropic $z-$direction.

Conductivity anisotropy is important in condensed matter physics, because the divergence of the resistivity anisotropy, and the temperature-linear resistivity at optimal doping of the cuprates are among the most puzzled problems to theorists. Normal-state transport in high--$T_c$ cuprates has become one of the most challenging topics in condensed matter physics. A clear understanding of the normal-state transport properties of cuprates is considered as a key step towards understanding the pairing mechanism for high- temperature superconductivity.

To calculate the DC conductivity along the anisotropic direction, we must impose a constant electric field in the $z$ direction with magnitude $E_z$.
Now, we  consider a small perturbation along the $z$-direction
\bea
&&A_{z}=-E_z t+\delta a_{z} (r), ~~~g_{t z}=\delta g_{t z}(r),\nonumber\\&&~~~g_{r x_i}=g_{zz}\delta h_{r z}(r),~~~\psi=k_z z+\delta \chi_z.
\eea
In this case, the conserved current is given by
\be
J^z=-\sqrt{-g}g^{rr}g^{zz}Z(\phi)\partial_r a_z+\delta g_{tz} g^{zz}q.
\ee
The DC conductivity along the anisotropic direction is
\be
\sigma_{zz}=\bigg(g^{\frac{d}{2}-1}_{xx}g^{-\frac{1}{2}}_{zz}Z_H+\frac{ q^2}{k^2_z Y_H g^{\frac{d-2}{2}}_{xx}\sqrt{g_{zz}}}\bigg)\bigg|_{r=\rh}.
\ee
Therefore, the determinant of the DC electric conductivity at zero charge density is then obtained as
 \be
\prod_{i} {\sigma_{ii}}\bigg|_{q=0}=g^{(d-1)(d-3)/2}_{xx}g^{(d-3)/2}_{zz}Z^{d-1}_H.
\ee
Comparing with the area density $\mathcal{A}=(g^{d-2}_{xx}g_{zz})^{1/2}$, we can reproduce the relation given in (\ref{scaling}).  In the next section, we will give several examples on
the effectiveness and universality of the relation (\ref{scaling}), in particular, the case with more than one gauge fields.

The ratio between the isotropic and anisotropic DC conductivities in the zero- charge limit $q\rightarrow 0$ can be easily evaluated
$ \frac{\sigma_{zz}}{\sigma_{jj}}=\frac{g_{xx}}{g_{zz}}\bigg|_{r=\rh}. $
For notably small $k_j \sim k_z \ll q$, the translation symmetry is weakly broken and it  is expected to have a Drude peak of conductivity.
The dissipation term becomes dominant when the particle-hole production term is negligible. The conductive anisotropy becomes
$ \label{ansicon} \frac{\sigma_{zz}}{\sigma_{jj}} \sim \frac{ \sqrt{g_{xx}}}{ \sqrt{g_{zz}}}\bigg|_{r=\rh}.
$
A concrete calculation was provided in Ref.\cite{gl}, where the DC conductivity along the $z-$direction  exhibits an insulating behavior with $d \rho_{DC}/d T< 0$, which is consistent with the $c-$axis transport behavior for underdoped cuprates.
The R-charged version of the anisotropic black brane solution was constructed  using the non-linear Kaluza-Klein reduction of type--IIB supergravity \cite{ge,ge1}, where $g_{x_i x_i}=\rh^2 e^{-\phi_H/2}$ and $g_{zz}=\rh^2 e^{-3\phi_H/2}$. Therefore, we achieve the result $\sigma_{zz}/\sigma_{ii}=e^{\phi_H} < 1$ since $\phi_H>0$.

This result is qualitatively  consistent with our observation in cuprates,  where the resistive anisotropy $\rho_c/\rho_{ab}$  varies between from $10$ to over $10^6$ at the critical temperature \cite{brooks}.
The ratio of shear viscosities also satisfies the above relation: $\eta_{x_i x_j, x_i x_j}/\eta_{x_i z,x_i z}={\sigma_{zz}}/{\sigma_{jj}}$.

\section{3. Examples}
{\it   Example 1.---} As a toy model, the simplest case of linear-T resistivity can be derived from the $2+1$-dimensional charged BTZ black holes. Utilizing the action (\ref{uniaction}), we simply set
\be
d=2,~~~\phi=0,~~~V(\phi)=2,~~~Y(\phi)=1,~~~Z(\phi)=1.
\ee
One of the solutions to the above action is given by the $2+1$-dimensional charged BTZ metric and a linear scalar field
\bea
&&ds^2=\frac{1}{z^2}\bigg(-f(z)dt^2+dx^2+\frac{dz^2}{f(z)}\bigg),\\
&&f(z)=1-\frac{z^2}{z^2_h}+\frac{\mu^2+c^2_1}{2}z^2\ln\frac{z}{z_h},\\
&&A_t(z)=\mu \ln\frac{z}{z_h},~~~\psi_1=c_1 x.
\eea
The black hole temperature  and entropy density are given by  $ T= \frac{4-z^2_h(\mu^2+c^2_1)}{8\pi z_h}$ and $s=4\pi z^{-1}_h \sim T$, respectively. In the higher temperature limit, the DC conductivity is obtained as
\be\label{btzs}
\sigma_{xx}=\frac{1}{T}+\frac{ \mu^2}{c^2_1 T}.
\ee
The universal relation (\ref{scaling}) is simply obeyed by this $2+1$-dimensional black hole.
In addition, the linear-T resistivity appears both in the quantum critical term and momentum dissipation term.
In the zero charge limit, we arrive at $\sigma_{QC}=1/T$. Physics in $1+1$-dimensions involve very interesting phenomena in condensed matter physics such as spin chains, quantum wires and Luttinger liquids. The first term in equation (\ref{btzs}) is qualitatively consistent with the result obtained in Ref. \cite{anderson}. Note that the shear viscosity bound is not violated by the BTZ black hole considered here. In the following, we will consider shear viscosity violated but the electric conductivity formula satisfied examples.

 {\it   Example 2.---}To demonstrate how the linear resistivity can be realized in higher dimensions, we consider an anisotropic systems with the following action:
\bea
S &=&\int d^5 x \sqrt{-g}\bigg[R+12 \Lambda -\frac{1}{2}(\partial \phi)^2-\frac{1}{2}e^{2\alpha\phi} (\partial \chi_3)^2\nonumber\\&-&\frac{1}{4}F^2-\frac{1}{2}\sum^{2}_{i=1}(\partial \chi_i)^2\bigg].
\eea
In the low-temperature limit, the solution is given by
\bea
&&ds^2=l^2\bigg(-r^2 f(r)dt^2+\frac{d r^2}{r^2 f(r)}+r^2 dx^2+r^2 dy^2\nonumber\\&&+ r^{\frac{4 \alpha^2}{1+2\alpha^2}}dz^2\bigg),\\
&&\chi_i=\beta_{ia}x^a, ~\chi_3= c_3 z, ~~\phi=\frac{2\alpha}{1+2\alpha^2}\ln \frac{\rh}{r},\nonumber\\&& ~~c_3=\frac{\sqrt{2(3+8\alpha^2)}}{1+2\alpha^2},~~l^2=\frac{3+8\alpha^2}{4+8\alpha^2}.
\eea
It is easy to verify that the above ansatz yields a solution in the absence of charge
\bea
&&f(r)=1-\frac{\beta^2(1+2\alpha^2)r^2}{(2+8\alpha^2)\rh^2}+\bigg[\frac{\beta^2(1+2\alpha^2)}{ (2+8\alpha^2)r^2}-1\bigg]\bigg(\frac{\rh}{r}\bigg)^{\frac{l^2}{4}},\nonumber\\
&&\beta^2=\frac{1}{2}\sum^2_1 \vec{\beta}_a \cdot \vec{\beta}_a, ~~\vec{\beta}_a \cdot \vec{\beta}_b=\beta^2 \delta_{ab}.
\eea
The Hawking temperature is given by $T=\frac{\rh l^2}{16 \pi}-\frac{\beta^2(1+2\alpha^2)l^2}{(2+8\alpha^2)\rh} $ and the entropy density $s=4\pi \rh^{2+\frac{2\alpha^2}{1+2\alpha^2}}$. The quantum critical conductivity at high temperature is given by
\bea
\sigma_{jj}=\rh^{\frac{2\alpha^2}{1+2\alpha^2}},~~~
\sigma_{zz}= \rh^{2-\frac{2\alpha^2}{1+2\alpha^2}}.
\eea
One can easily verify that the relation (\ref{scaling}) is valid for this black hole background.
For the special case of $\alpha^2=-\frac{1}{4}$ and $\beta^2/\rh\rightarrow 0$ but $\beta^2/(1+4\alpha^2)$ is finite, we can  easily obtain
\bea
\sigma_{jj}=\frac{16 \pi  }{l^2} T^{-1} ,~~~\sigma_{zz}= \frac{4096 \pi^3}{l^6} T^3.
\eea
The resistivity, i.e., $\sigma^{-1}_{jj}$, varies linearly  with $T$, which provides a phenomenological account for the linear resistivity of strange metals. $\sigma_{zz}$  monotonically decreases with decreasing temperature
and behaves as $T^3$. The metallic behavior in the ``ab"-plane and insulating behavior along the ``c"-axis indicate that this anisotropic model captures the key features of the high- $T_c$ transport properties in the normal state. The experimental data of $\sigma_c$  is proportional to $T^3$ for $\rm YBa_2 Cu_3 O_{6.95}$ \cite{hardy}, which is consistent with  our result   that the $T^3$ power law of the c-axis conductivity  \cite{xiang} is  notably \textit{ universal} for cuprates.

As argued in the previous section, for systems with rotational--symmetry breaking, vector  Goldstone bosons are generated. Such Goldstone modes correspond to the broken Lorentz symmetry  in the boundary theory. From the Kaluza-Klein reduction, we know that the off--diagonal components of the metric, whose perturbations carry spin 1, induce gauge fields in the dimensionally reduced
theory.  Jain et al. argued that the conductivity of these gauge fields, is proportional to the spin- 1 viscosity components $\eta_{x_i z,x_i z}$ \cite{jain}, which  motivates us to conjecture that the dc electrical conductivities
obeys a universal lower bound in anisotropic systems.

{\it   Example 3.---} It would be interesting to consider black holes with a hyperscaling violating factor.
 we consider the following action
\bea\label{action}
  S&=&\int\md^{4}x\sqrt{-g}\big[{R}-\frac{1}{2}(\partial\phi)^2
  -\frac{1}{4}\sum_{i=1}^{2}\me^{\lambda_{i}\phi}(F_{i})^{2}\nonumber\\&
  -&\frac{1}{2}\me^{\eta\phi}\sum_{i=1}^{2}(\partial\chi_{i})^{2}
  +\sum_{i=1}^{2}V_{i}\me^{\gamma_{i}\phi}\big],
\eea
where $\lambda_{i},\eta,\gamma_{i},V_{i}$ are undetermined constant parameters and $Z_i(\phi)=\me^{\lambda_{i}\phi}$, $Y(\phi)=\me^{\eta\phi}$ and $V=V_{i}\me^{\gamma_{i}\phi}$.
Note that there are two $U(1)$ gauge $F^{(1)}_{rt}$ and $F^{(2)}_{rt}$ in which the first gauge field plays the role of an auxiliary field, making the geometry asymptotic Lifshitz, and the second gauge field is the real Maxwell field.
 We obtain one of the black hole solutions as
 \bea
 ds^2 &=& r^{-\theta}\bigg(-r^{2z}f(r)dt^2+\frac{dr^2}{r^2 f(r)}+r^2 dx^2+r^2 dy^2\bigg),\nonumber\\
 F^{(1)}_{rt} &=& r^{z-\theta+1}q_1,~~~  F^{(2)}_{rt} =r^{\theta-z-1}q_2,\nonumber\\
 \phi &=& \nu \ln r=\sqrt{(\theta-2)(\theta-2z+2)},~~~ \chi_i=\beta x_i.
  \eea
  The parameters are solved and take the following values
  \bea
   \lambda_1&=&\frac{\theta-4}{\nu}, ~~~\lambda_2=\eta=\frac{\nu}{\theta-2},~~~\gamma_1=\frac{\theta}{\nu},\nonumber\\
   \gamma_2&=&\frac{\theta+2z-6}{\nu},~V_1=\frac{(z-\theta+1)q^2_1}{2(z-1)},\\ V_2&=&\frac{q^2_2(2z-\theta-2)}{4(z-2)}. \nonumber
  \eea
  The blacken function $f(r)$ yields the form
  \be
  f(r)=1-m r^{\theta-2-z}-\beta^2 r^{\theta-2z}+\frac{q_2 r^{2z-6}}{4(z-2)(3z-\theta-4)}.\nonumber
  \ee
  The black hole solution presented here is very general and for special value of $\theta$ and $z$ (i.e. $\theta=0$ and $z=1$), one recovers the Reissner-Nordstr$\ddot{o}$m-AdS black hole solution with linear scalar fields.
 The event horizon locates at $f(\rh)=0$. The Hawking temperature is given by
  \be
  T=\frac{1}{4\pi}\left[(z-\theta+2)\rh^z+\frac{\beta^2 \rh^{\theta-z}}{\theta-2}+\frac{q^2_2 \rh^{3z-6}}{4(z-2)}\right].
  \ee
  Similar black hole solution has been obtained in \cite{ge16} and a new computational tool for computing the DC transport coefficients was presented in \cite{ge16} and \cite{ge1606}.
  Now we have two $U(1)$ gauge fields, in general, there will be mixing terms between fluctuations of two gauge fields in the expressions of conductivities as discussed in \cite{ge1606}.
  The general conductivity matrix takes the form
  \bea \label{conmatrix}
  \sigma^{(11)}_{xx}&=&\rh^{\theta-4}+\frac{q^2_1 \rh^{2z-4}}{\beta^2}\rh^{2z-4},~\sigma^{(12)}_{xx}=\frac{q_1 q_2 \rh^{2z-4}}{\beta^2},\nonumber\\
  \sigma^{(22)}_{xx}&=&\rh^{\theta-2z+2}+\frac{q^2_2 \rh^{2z-4}}{\beta^2}\rh^{2z-4},~\sigma^{(21)}_{xx}=\sigma^{(12)}_{xx}.
  \eea
  Recall that  $\sigma^{(11)}_{xx}$ and $\sigma^{(22)}_{xx}$ are generated by two distinct electric fields, which are oriented along the $x-$direction. Similar conductivity matrix can be obtained along the $y-$direction. Substituting equation (\ref{conmatrix}) at zero charge density, the area density and gauge coupling $Z_i$ into (\ref{scaling}) and evaluating at the event horizon, we can see that (\ref{scaling}) is satisfied for general value of $\theta$ and $z$.

   As a side note and just for simplicity, we set $z=1$ from the beginning, so that the first gauge field and $q_1$ vanish in the action.
  Since this is an isotropic systems, followed the DC conductivity formula given in (\ref{isocon}), the DC conductivity reads
  \be \label{sigii}
  \sigma_{ii}=\rh^{\theta}+\frac{q^2_2}{\beta^2}\rh^{-2}.
  \ee
  One can easily verify that the general relation (\ref{scaling}) is satisfied by (\ref{sigii}) when $q_2=0$ at any physical temperature.
 Moreover, the linear- and quadratic- in temperature resistivity can be realized when $z=1$ and $\theta=-1$ in the limit $\rh \sim T$. That is to say
  \be
  \sigma_{ii} \sim \frac{1}{T}+\frac{q^2_2}{\beta^2}\frac{1}{T^2}.
  \ee
    This matches with the experimental observations of cuprates:  The in-plane resistivity varies approximately linearly with temperature at high temperature, while as temperature cools down, the resistivity is governed by the Fermi-liquid-like $T^2$ behavior \cite{donos3,ge16}.

  As demonstrated in \cite{wang} and \cite{ling}, the shear viscosity in this $4$-dimensional isotropic systems will violate the KSS conjecture because of the momentum dissipation and the appearance of the effective mass of gravitons. However, the relation (\ref{scaling}) stills remain to be uninfluenced  even in the presence of the translational symmetry breaking.

  We can also prove that for the electric conductivity bound \cite{schalm} violated system considered in \cite{matteo2,weijia}, the universal formula of DC electrical conductivity (\ref{scaling}) is still valid. In this sense,  the formula (\ref{scaling}) seems to be more universal than the shear viscosity bound \cite{kss} and the electric bound \cite{schalm}.

\section{4. Discussions and conclusions}
Considering the above facts, we propose that the determinant of the quantum critical conductivity matrix has a scaling relation with the black hole horizon area: $\prod_i \sigma_{ii}|_{q_i=0}=\mathcal{A}^{d-3} \prod_i Z_i^{d-1}(\rh) $.


In order  to recast (\ref{scaling}) in a experimental testable manner, we consider a generalized spatially anisotropic black hole with following line-elements
\be
ds^2=- r^2 f(r) dt^2+\frac{dr^2}{r^2 f(r)}+r^2 \sum^p_{i=1}dx^2_i+r^{\frac{2}{z}}\sum^{d-1}_{j=p+1}dy^2_j,
\ee
where $f(r)$ is the blacken factor with event horizon locates at $r=\rh$. The horizon area density $\mathcal{A}=\rh^{p+(d-1-p)/z}\sim T ^{p+(d-1-p)/z}$, where we have assumed $\rh\sim T$.
Therefore, the conductivity formula reduces to
\be \label{Tbound}
\prod_i \sigma_{ii}|_{q_i=0}= \mathcal{C} T ^{\big(p+\frac{(d-1-p)}{z}\big)(d-3)} \prod_i Z_i^{d-1}(\rh),
\ee
where $\mathcal{C}$ is a specific constant to be determined. Equation (\ref{Tbound}) is a generalized form of quantum critical conductivity and it can recover the quantum critical conductivity given in Refs.\cite{damle,sachdev,liu} under the conditions $Z_H=1$ and $p=0$. Note that we do not consider the hyperscaling violating factor $\theta$ here.

In the slow relaxation limit, the DC conductivity can be written as the sum of an explicit charge--dependent term  and a quantum critical term
$ \sigma^{DC}_{ii}=\sigma^{QC}_{ii}+\frac{q^2}{\varepsilon+p}\tau^{L}_{ii}, $
where $\varepsilon$ and $p$ are the energy density and pressure, respectively, and $\tau_{L}$ is a time scale associated with the impurity/lattice.
Therefore, this theory has a ``universal" finite conductivity even without a net charge density. The quantum critical current that is carried by the particle-hole pairs of opposite momenta, controls the rate
at which charge diffuses instead of the momentum relaxation. However, in an anisotropic system, the quantum critical conductivity $\sigma_{QC}$ along different directions should not be identical.

In this study, we propose a universal formula of the DC electrical conductivity applied to both anisotropic and isotropic systems. Holography provides us a uniquely tractable method to study those strongly interacting systems without quasiparticles. In incoherent metals without a Drude peak, the transport is described by diffusive physics in terms of the diffusion of charge and energy instead of momentum diffusion. Thus, one can propose that the conductivity is a more universal physical quantity in such systems. In the absence of isotropy, different metric perturbations break up into components with different spin values. The shear viscosity in a rotationally invariant field theory is proportional
to the graviton absorption via $\eta=\Sigma(0)/2\kappa^2$. The spin 2 metric perturbation component obeys the equation of motion  of a minimally coupled massless scalar. Therefore, the absorption cross-section of a graviton is equivalent to that of a scalar field. A theorem on the scalar absorption cross-section states that in the larger-wavelength limit, $\mathcal{A}=\Sigma(0)$. So, the shear viscosity is proportional to the black hole entropy density because $s=\mathcal{A}/4G$. However, for rotational-symmetry-breaking systems, for a metric perturbation with spin 1, the equation of motion for $h_{z}^{x_i}$ cannot be written in the form  of a minimal coupled massless scalar. Instead, its equation of motion can be recast in a  similar  form to the Maxwell equation: $\nabla_{\mu} f^{\mu\nu}+\nabla_{\mu}g_{xx} f^{\mu\nu}/g_{xx}=0$. Even for isotropic but translation invariance broken systems, an effective mass term is generated in the equation of motion of the tensor mode so that the shear viscosity is not equivalent to the black hole entropy density. Based on this formula (\ref{scaling}), we however are able to provide examples yielding the linear temperature resistivity and insulating behavior are realized in the isotropic plane and out-of-plane, respectively. We are also able to
 realize both the linear-T and quadratic-T contributions to the resistivity by constructing a $4-$dimensional isotropic black hole solution.

The DC electrical conductivity formula proposed here (i.e., (\ref{scaling}) and (\ref{Tbound})) can provide some insights on future model building of linear temperature resistivity in holographic theory and it is experimentally testable because it can recover the form in  Ref. \cite{sachdev}. If the dual Maxwell theory must be minimally coupled, the presence of  the linear temperature resistivity and the universality of the conductivity bound infers that anisotropy is a fundamental factor to be considered in normal-state of high temperature superconductors. We expect that our result is falsifiable in the future study.\\
\textit{Acknowledgements.---} We thank R. G. Cai, Sean Hartnoll, Yi Ling and R. Samanta for the useful discussions.
XHG was partially supported by NSFC,
China (No.11375110) and the Grant (No. 14DZ2260700) from  Shanghai Key Laboratory of High Temperature Superconductors. SJS was supported by the NRF, Korea (NRF-2013R1A2A2A05004846). SFW was supported partially by NSFC
China ( No. 11275120 and No. 11675097).

\end{document}